# A probabilistic deep learning approach to automate the interpretation of multi-phase diffraction spectra

Nathan J. Szymanski[1,2], Christopher J. Bartel[1,2], Yan Zeng[2], Qingsong Tu[2], and Gerbrand Ceder[1,2*]

**Abstract**

Autonomous synthesis and characterization of inorganic materials requires the automatic and accurate analysis of X-ray diffraction spectra. For this task, we designed a probabilistic deep learning algorithm to identify complex multi-phase mixtures. At the core of this algorithm lies an ensemble convolutional neural network trained on simulated diffraction spectra, which are systematically augmented with physics-informed perturbations to account for artifacts that can arise during experimental sample preparation and synthesis. Larger perturbations associated with off-stoichiometry are also captured by supplementing the training set with hypothetical solid solutions. Spectra containing mixtures of materials are analyzed with a newly developed branching algorithm that utilizes the probabilistic nature of the neural network to explore suspected mixtures and identify the set of phases that maximize confidence in the prediction. Our model is benchmarked on simulated and experimentally measured diffraction spectra, showing exceptional performance with accuracies exceeding those given by previously reported methods based on profile matching and deep learning. We envision that the algorithm presented here may be integrated in experimental workflows to facilitate the high-throughput and autonomous discovery of inorganic materials.

[1]Department of Materials Science & Engineering, UC Berkeley, Berkeley, CA 94720, USA

[2]Materials Sciences Division, Lawrence Berkeley National Laboratory, Berkeley, CA 94720, USA

*correspondence to gceder@berkeley.edu

## Introduction

The development of high-throughput and automated experimentation has ignited rapid growth in the amount of data available for materials science and chemistry[1,2]. Unlocking the physical implications of resulting datasets, however, requires detailed analyses that are traditionally conducted by human experts. In the synthesis of inorganic materials, this often entails the manual interpretation of X-ray diffraction (XRD) spectra to identify the phases present in each sample. Past attempts to automate this procedure using peak indexing[3,4] and full profile matching[5,6] algorithms have been limited by modest accuracy, in large part because measured spectra usually deviate from their ideal reference patterns (e.g., due to defects or impurities). Consequently, the analysis of XRD spectra widely remains a manual task, impeding rapid materials discovery and design. To alleviate this bottleneck, deep learning based on convolutional neural networks (CNNs) has recently emerged as a potential tool for automating the interpretation of diffraction spectra with improved speed and accuracy[7,8].

Previous work has demonstrated that CNNs can be used to perform symmetry classification[9-11] and phase identification[12,13] from XRD spectra of single-phase samples. Given the lack of well-curated diffraction data obtained experimentally, training is most commonly performed on labeled sets of *simulated* spectra derived from known crystalline materials, e.g., in the Inorganic Crystal Structure Database (ICSD)[14]. However, because many factors can cause cause differences between observed and simulated diffraction peaks, this approach can be problematic for extension to experimentally measured XRD spectra. Vecsei *et al.* demonstrated that a neural network trained on simulated spectra produced an accuracy of only 54% for the classification of experimentally measured diffraction spectra extracted from the RRUFF database[10]. To overcome this limitation, simulated spectra can be augmented with perturbations designed to emulate possible artifacts. For example, Oviedo *et al.* trained a CNN using simulated spectra augmented with random changes in their peak positions and intensities, which were chosen to account for texture and epitaxial strain in the thin films being studied. The resulting model correctly classified the space group for 84% of diffraction spectra measured from 115 metal halide samples[7]. We propose that generalization of existing methods to handle complex XRD spectra requires a more complete data augmentation procedure that properly accounts for all the artifacts that frequently arise during sample preparation and synthesis.

To extend the application of CNNs to mixtures of materials, Lee *et al.* constructed a training set of multi-phase spectra that were simulated using linear combinations of single-phase diffraction spectra from 38 phases in the quaternary Sr-Li-Al-O space[8]. Their model performed well in the identification of high-purity samples, with 98% of all phases correctly labeled based on 100 three-phase spectra. However, the combinatorial nature of their technique requires an exceptionally high number of training samples (nearly two million spectra from 38 phases), which restricts the inclusion of experimental artifacts *via* data augmentation. Moreover, because the number of training samples increases exponentially with the number of reference phases, the breadth of the composition space that can be efficiently considered is limited. Proposing an alternative approach, Maffettone *et al.* designed an ensemble model trained on simulated single-phase spectra to yield a probability distribution of suspected phases for a given spectrum[12]. From this distribution, the authors infer that high probabilities suggest that the corresponding phases are present in the mixture. While this method avoids combinatorial explosion and thus allows many experimental artifacts to be included during training, it sometimes leads to confusion as obtaining comparable probabilities for two phases does not necessarily imply that both are present. Rather, it may simply mean that the algorithm has difficulty distinguishing between the two phases. An improved treatment of multi-phase spectra therefore necessitates an approach that (i) allows artifacts to be incorporated across many phases and (ii) distinguishes between probabilities associated with mixtures of phases as opposed to similarities between single-phase reference spectra.

In this work, we introduce a novel deep learning technique to automate the identification of inorganic materials from XRD spectra of single- and multi-phase samples. In our approach, training spectra are generated with physics-informed data augmentation whereby experimental artifacts (strain, texture, and domain size) are used to perturb diffraction peaks. The training set is built not only from experimentally reported stoichiometric phases, but also from hypothetical solid solutions that account for potential off-stoichiometries. An ensemble CNN is trained to yield a distribution of probabilities associated with suspected phases, which is shown to be a surrogate for prediction confidence. We extend this probabilistic model to the analysis of multi-phase mixtures by developing an intelligent branching algorithm that iterates between phase identification and profile subtraction to maximize the probability over all phases in the predicted mixture. To demonstrate the effectiveness of our CNN, training and testing were conducted using diffraction spectra derived from materials in the broad Li-Mn-Ti-O-F composition space given their structural diversity and technological relevance (e.g., for Mn-based battery cathodes)[15]. By also systematically testing on a dataset of experimentally measured XRD spectra designed to sample complexities that often arise during synthesis, we show that our algorithm achieves considerably higher accuracy than state-of-the-art profile matching techniques as well as previously developed deep learning-based methods.

## Methods

### Stoichiometric reference phases

The identification of inorganic materials from their XRD spectra relies on the availability of suitable reference phases that can be compared to samples of interest. In this work, we focus on the Li-Mn-Ti-O-F chemical space (and subspaces) and retrieved all 1,216 corresponding entries from the ICSD[14]. For the identification of stoichiometric materials, we excluded 386 entries with partial occupancies from this set. To remove duplicate structures from the remaining 830 entries, all unique structural frameworks were identified using the pymatgen structure matcher[16]. For each set of duplicates, the entry measured most recently at conditions nearest ambient (20 °C and 1 atm) were retained. Based on these selection criteria, 140 unique stoichiometric materials listed in **Supplementary Table S1** were tabulated and used as reference phases.

### Non-stoichiometric reference phases

Although many solid solutions are available in the ICSD, they generally cover a narrow composition range while leaving others sparse. We therefore designed an algorithm to extend the space of non-stoichiometric reference phases by using empirical rules to construct hypothetical solid solutions between the available stoichiometric materials. To determine which phases may be soluble with one another, all combinations of the 140 stoichiometric references phases in the Li-Mn-Ti-O-F space were enumerated and two criteria were considered for each pair. First, solubility requires that the two phases adopt similar structural frameworks, which was verified using the pymatgen structure matcher[16]. Second, based on the Hume-Rothery rules[17], the size mismatch between any ions being substituted with one another should be $\leq 15\%$. To estimate the ionic radii of all species comprising each phase, oxidation states were assigned using the composition-based oxidation state prediction tool in pymatgen[16]. In cases where mixed oxidation states are present (e.g., $Mn^{3+/4+}$), we chose to focus on the state(s) that minimizes the difference between the radii of the ions being substituted and therefore increases the likelihood for solubility. As will be shown by our test results, including more reference phases does not lead to a substantial decrease in accuracy; hence, it is preferable to overestimate solubility such that more structures are created as potential references.

Based on the 140 stoichiometric reference phases in the Li-Mn-Ti-O-F space, 43 pairs of phases were found to satisfy both solubility criteria described above. The phases in each pair were treated as end-members, from which interpolation was used to generate a uniform grid of three intermediate solid solution compositions. For example, between spinel $LiMn_2O_4$ and $LiTi_2O_4$, intermediate compositions take the form $LiMn_{2-x}Ti_xO_4$ with $x \in \{0.5, 1.0, 1.5\}$. The lattice parameters of hypothetical solid solutions were linearly interpolated between those of the corresponding end-members in accordance with Vegard's law[18]. Atomic positions and site occupancies were similarly obtained by interpolating between equivalent sites in the end-members. This procedure gave a total of 129 hypothetical solid solution states from the 43 pairs of soluble phases. Excluding 14 duplicates resulted in 115 distinct solid solutions, listed in **Supplementary Table S2**. The code for generating hypothetical solid solutions for an arbitrary group of reference phases is available at https://github.com/njszym/XRD-AutoAnalyzer.

## Data augmentation

From the reference phases in the Li-Mn-Ti-O-F space, we built an augmented dataset of simulated XRD spectra with the goal of accurately representing experimentally measured diffraction data. Physics-informed data augmentation was applied to produce spectra that sample possible changes in peak positions, intensities, and widths. Shifts in peak positions ($2\theta$) were derived using strain tensors that preserve the space group of the structure. Modified unit cells were created with up to $\pm 4\%$ strain applied to each lattice parameter. Peak widths were broadened by simulating domain sizes ranging from 1 nm (broad) to 100 nm (narrow) through the Scherrer equation[19]. Peak intensities were varied to mimic texture along preferred crystallographic planes. This was done by performing scalar products between the peak indices and randomly selected Miller indices $(hkl)$, followed by a normalization that scaled peak intensities by as much as $\pm 50\%$ of their initial values. The bounds chosen here are designed to reflect the range of artifacts that can occur during inorganic synthesis. We note that larger variations may arise when substantial off-stoichiometry is present; however, this situation was treated separately by the addition of non-stoichiometric solid solutions as reference phases. In **Fig. 1a**, we illustrate the effect of each of the three experimental artifacts on the XRD spectrum of spinel $Mn_3O_4$ as an example. Each artifact was applied separately to the simulated spectrum by taking 50 random samples from a normal distribution (e.g., between −5% and +5%), resulting in 150 augmented spectra per reference phase (50 samples for each of the three artifacts). Applying this procedure to all 255 references phases, including both experimentally reported stoichiometric materials and hypothetical solid solutions, resulted in 38,250 simulated diffraction spectra. Further details regarding data augmentation and spectrum simulation are provided in **Supplementary Note 1**. The code for performing data augmentation for an arbitrary group of reference phases is available at https://github.com/njszym/XRD-AutoAnalyzer.

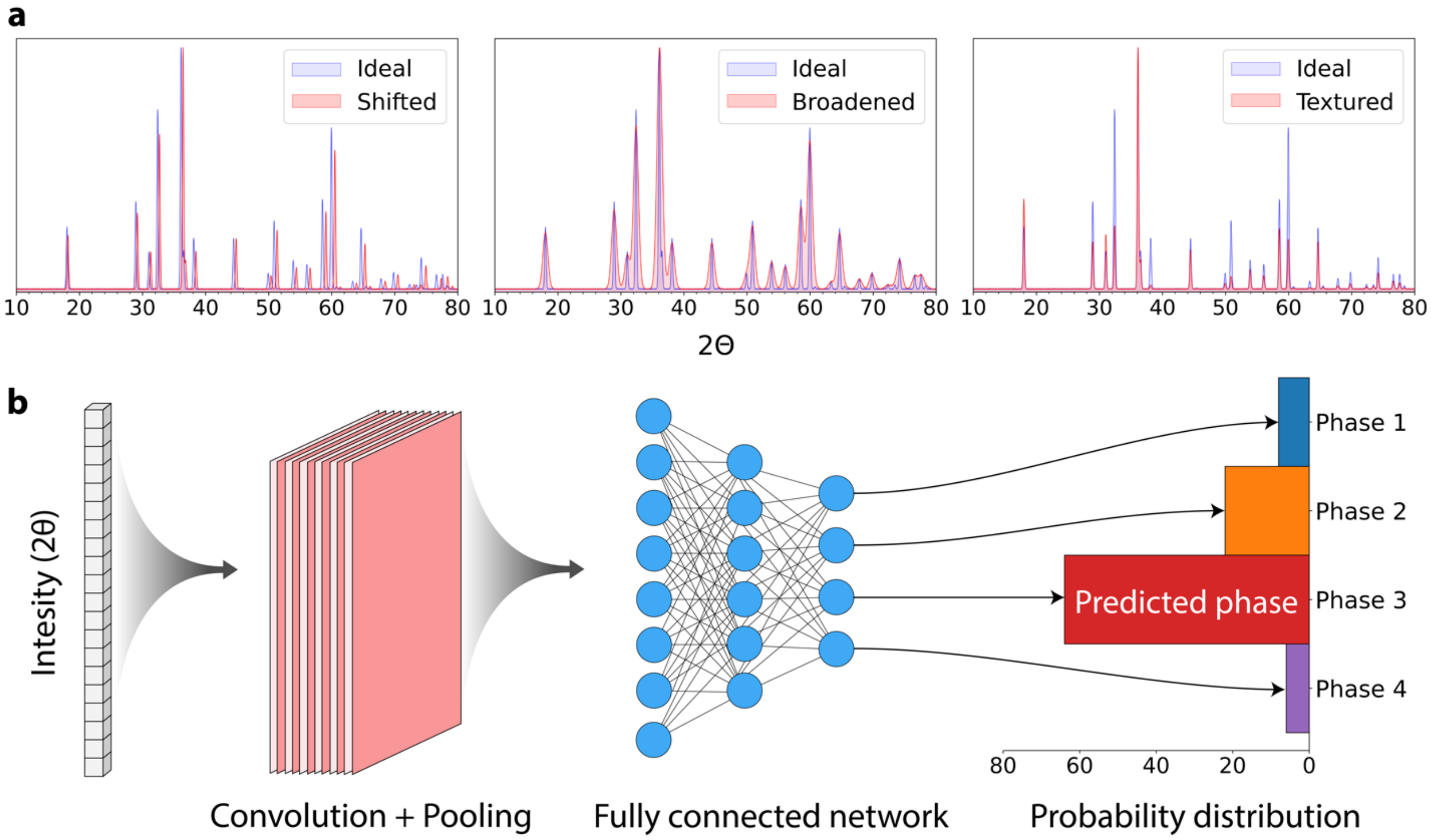

**Figure 1: (a)** An illustration of the data augmentation procedure designed to sample possible experimental artifacts including peak shift associated with cell strain, peak broadening related to small domain size, and peak intensity variation caused by texture. **(b)** A schematic of the deep learning pipeline used to map XRD spectra onto a probability distribution of suspected phases.

## Convolutional neural network

The workflow used to classify a given XRD spectrum is displayed in **Fig. 1b**. Similar to previous work[8], diffraction spectra are treated as one-dimensional vectors that contain 4,501 values for intensity as a function of $2\theta$. The range of $2\theta$ is set from $10°$ to $80°$, which is commonly used for scans with Cu $K_\alpha$ radiation ($\lambda = 1.5406$ Å). The intensities (represented as 4,501-valued vectors) serve as input to a CNN that consists of six convolutional layers, six pooling layers, and three fully connected layers. Training was carried out with five-fold cross-validation using 80% of the simulated diffraction spectra, with the remaining 20% reserved for testing (i.e., excluded from training and validation). Details regarding the architecture of the CNN and the hyperparameters used during training are given in **Supplementary Note 2**. The code used for training is also available at https://github.com/njszym/XRD-AutoAnalyzer. To classify spectra outside of the training set, an ensemble approach was used whereby 1,000 individual predictions are made with 60% of connections between the fully connected layers randomly excluded (i.e., using dropout) during each iteration. The probability that a given phase represents the spectrum is then defined as the fraction of the 1,000 iterations where it is predicted by the CNN. The resulting distribution may be treated as a ranking of suspected phases in the sample, with corresponding probabilities providing measures of confidence.

## Intelligent branching algorithm

Given that the CNN was trained only on single-phase XRD spectra, additional methods were developed to automate the identification of materials in multi-phase mixtures. In our workflow, we use an iterative procedure where phase identification is followed by profile fitting and subtraction. Once a phase is identified by the CNN, its diffraction peaks are simulated and fit to the spectrum in question using dynamic time warping (DTW), a well-known technique for correlating features in time series[20]. The resulting profile of the identified phase is then subtracted to produce a modified spectrum that is representative of the mixture minus the phase that has already been identified. This process is repeated until all significant peaks are attributed to a reference phase; i.e., the cycle is halted once all intensities fall below 5% of the initially measured maximum intensity. Further details regarding the techniques used to perform profile fitting and subtraction are described in **Supplementary Note 3**, and the corresponding code is available at https://github.com/njszym/XRD-AutoAnalyzer.

Following the iterative procedure outlined above, one could identify a multi-phase mixture by using the collection of most probable phases given by the model at each step. However, because the spectrum is affected by all prior phases that have been identified, such a method over-prioritizes the first iteration of phase identification. In cases where the first phase predicted by the CNN is incorrect, the spectrum resulting from profile fitting and subtraction will contain diffraction peaks that do not accurately represent the remaining phases in the sample. All subsequent analyses will therefore be less likely to identify these phases. To improve upon this approach, we developed an intelligent branching algorithm that gives equal importance to each iteration of phase identification. In **Fig. 2**, we illustrate how the algorithm evaluates several possible sets of phases to classify a diffraction spectrum derived from a mixture of $Li_2TiO_3$, $Mn_3O_4$, and $Li_2O$. At each step, the CNN generates a list of suspected phases along with their associated probabilities. As opposed to considering only the most probable phase at each iteration, the branching algorithm investigates all phases with non-trivial probabilities ($\geq$ 10%). By following the spectrum associated with the subtraction of each suspected phase, a "tree" is constructed to describe all combinations of phases predicted by the model. Once each route has been fully exhausted, the branch with the highest average probability is chosen as the final set of predicted phases (e.g., the green phases highlighted in **Fig. 2**). In this way, the algorithm maximizes the likelihood that predictions are representative of *all* phases contained in the actual mixture, as opposed to over-prioritizing the first iteration of phase identification. We found that this is an essential feature to predict multi-phase spectra correctly.

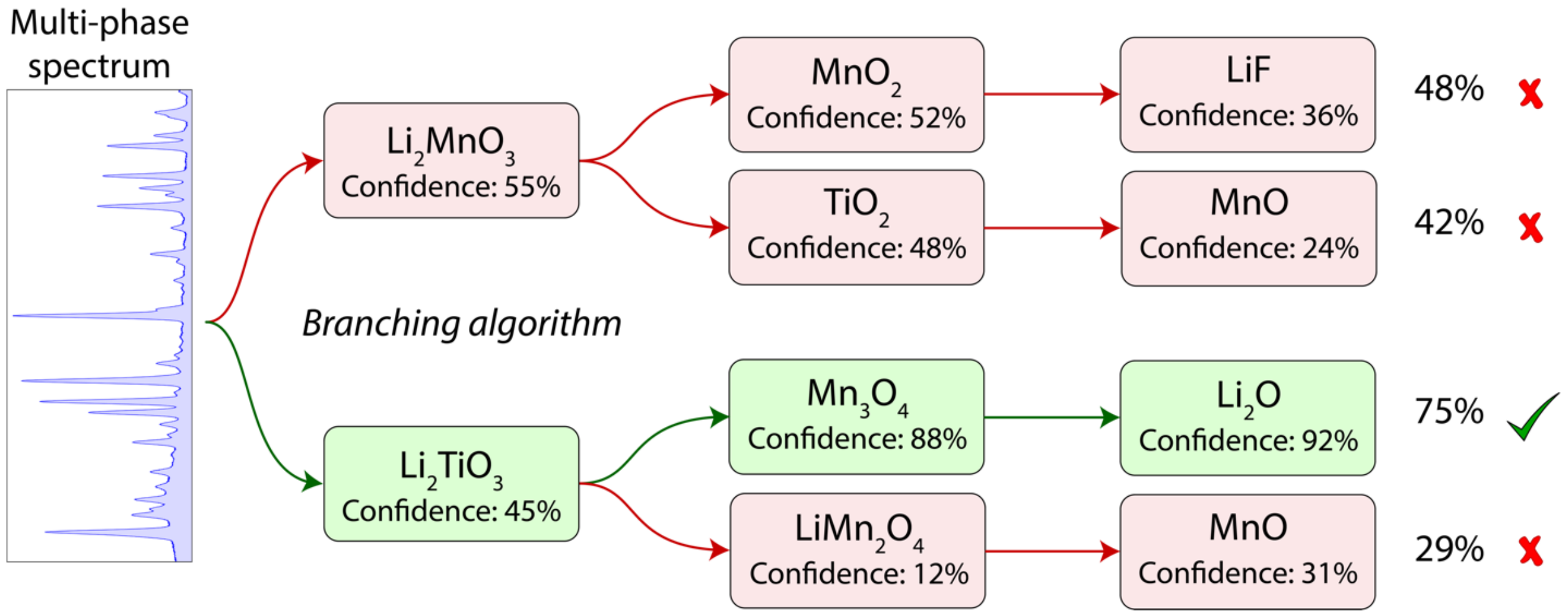

**Figure 2:** A schematic illustrating possible pathways enumerated by the branching algorithm for multi-phase identification. This method iteratively performs single-phase predictions followed by profile-stripping, at each step tabulating the probability associated with each phase. This process is repeated until all intensities fall below 5% of the original maximum value. From all branches developed, the one with the highest average probability (highlighted green above) across all levels is chosen as the most likely set of phases present in the mixture.

**Experimental measurements**

To further validate our model, we built an experimental dataset from a series of measurements designed to sample complexities that often arise during synthesis. Ten materials, listed in **Supplementary Note 4** with details regarding the experimental procedures, were chosen to span a range of structures and compositions in the Li-Mn-Ti-O-F space. For a benchmark on pristine single-phase spectra with no intended artifacts, we conducted precise diffraction measurements on each of the ten materials using carefully prepared, high-purity samples. The following modifications were then separately introduced such that each batch of samples contained one anticipated artifact: (i) samples were overlaid with Kapton tape during characterization to produce a diffuse background signal with a magnitude as large as 200% of the highest diffraction peak intensity; (ii) rapid scan rates (30°/minute) were used to generate noisy baseline signals with magnitudes reaching 5% of the maximum diffraction peak intensity; (iii) peak shifts as large as 0.4° were imposed by preparing thick pellets such that specimens were leveled slightly above the sample holder; (iv) broad peaks with full widths at half maxima as large as 1.5° were obtained by ball milling. Several additional materials were also made to sample changes in composition and site occupancy. Six samples of spinel $LiMnTiO_4$ were synthesized at temperatures of 900 °C, 950 °C, and 1000 °C followed by quenching or slow cooling based on previously reported procedures[21]. These samples were intended to contain differences in relative diffraction peak intensities owing to varied distributions of cation site occupancies. Non-stoichiometry was studied using four disordered rocksalt phases, each with a different composition made *via* solid-state synthesis. For the classification of multi-phase XRD spectra, ten two- and three-phase mixtures (listed in the **Supplementary Note 4**) were prepared from combinations of materials in the Li-Mn-Ti-O-F space that were chosen to include spectra with a substantial amount of peak overlap. The mixtures contained equal weight fractions of all constituent phases. To isolate the effects of multiple phases, these measurements were conducted on samples for which no experimental artifacts were purposefully incorporated.

## Results

### Identification of stoichiometric phases

As a first test case, we evaluated the performance of our model on simulated single-phase XRD spectra derived from the 140 stoichiometric reference phases in the Li-Mn-Ti-O-F space. Accordingly, the CNN was trained on 80% of the 21,000 generated spectra (140 materials × 150 augmentations) that were augmented to include physics-informed perturbations to their diffraction peak positions, widths, and intensities. The remaining 4,200 spectra were reserved for testing. To assess the ability of the CNN to handle artifacts not considered during training, the test set was also supplemented with spectra having diffuse and noisy background signals. A diffuse background was simulated by adding an XRD spectrum measured from amorphous silica to the diffraction peaks of the stoichiometric materials. Ten spectra were created for each phase (1,400 spectra total), with the maximum intensity produced by silica ranging from 100-300% of the maximum peak intensity of the reference phase. Another 1,400 spectra were simulated by adding Gaussian noise with magnitudes ranging from 1-5% of the maximum diffraction peak intensity. Before being passed to the CNN, these 2,800 spectra were pre-processed using the baseline correction and noise filtering algorithms described in **Supplementary Note 5**. This procedure is designed to replicate artifacts formed when imperfect corrections are made during pre-processing, which occasionally leads to the disappearance of minor peaks or leaves behind residual intensities related to amorphous impurities. Previous work has dealt with diffuse and noisy background signals by training on spectra with added baseline functions (e.g., polynomials)[9,12]. However, because these functions are randomly selected rather than derived from possible impurities or defects, they are unlikely to accurately represent experimental measurements[13]. With this in mind, our current approach relies only on physics-informed data augmentation to improve the match between simulated and experimentally measured spectra.

The performance of our model is compared to a known standard, the JADE software package from MDI[22]. JADE is a widely used program that can automate phase identification with conventional profile matching techniques[5]. During testing, JADE was employed without any manual intervention to ensure a consistent comparison with the CNN, as we are assessing the capability of our approach to perform phase identification as part of an autonomous platform. We emphasize that our model is not designed to replace manual techniques such as Rietveld refinement, but rather to provide more rapid and reliable predictions regarding phase identities. For this task, we applied both the trained CNN and JADE to the test set of simulated diffraction spectra that sample possible experimental artifacts *separately* as discussed in the **Methods**. In **Fig. 3a**, we compare the resulting accuracy of each method quantified as the fraction of phases correctly identified. Across the simulated test spectra, the CNN achieves a high accuracy of 94%. In contrast, JADE correctly identifies only 78% of phases when applied to the same set of spectra. To further verify the effectiveness of the CNN, an additional 1,400 spectra were simulated with *mixed* artifacts such that each spectrum contains all aforementioned perturbations to its diffraction peaks (shifting, broadening, and texture) as well as a diffuse and noisy background signal. This incorporates an additional level of complexity not included in the training set, where each spectrum contained just one type of perturbation. When applied to the new test set with mixed artifacts, the accuracy of the CNN decreases only 2% (from 94% to 92%), whereas the accuracy of JADE decreases 10% (from 78% to 68%).

The tests show promising results for the CNN, though its performance is not without error. We look to the underlying causes of the occasional misclassifications that occur by dividing the simulated test spectra into four major categories: those augmented *via* the individual application of peak shifts, peak broadening, peak intensity change, and background effects (including diffuse and noisy baselines). The training set remains unchanged from the previous paragraph. In **Fig. 3b**, we show the fraction of misclassifications that arise from each perturbation category. Of the 7,000 total test spectra, 418 are misclassified by the CNN. The largest portion (48%) of misclassifications occur for spectra containing peak shifts, which we attribute to the overlapping of diffraction peaks between similar phases. This most commonly occurs between isomorphic phases and, as a result, the CNN gives a higher accuracy for the identification of structure (96%) as opposed to composition (92%). We investigated the effects of increasing the bounds on strain that were used during training (beyond $\pm 4\%$); however, a decrease in accuracy was observed as larger strains were incorporated. For example, training on spectra derived from structures with strain as large as $\pm 6\%$ led to a lower accuracy of 86% when applied to the test set containing spectra with as much as $\pm 4\%$ strain. More details regarding the effects of strain are illustrated in **Fig. S1**. Relative to peak shifts caused by strain, spectra with broad peaks lead to fewer misclassifications, comprising 27% of errors. For this effect, misclassification occurs more frequently in low-symmetry structures as they contain many diffraction peaks that tend to overlap with one another upon broadening. Of the 113 spectra that are incorrectly classified by the CNN due to peak broadening, 82 are from phases with monoclinic or triclinic symmetry. The remaining artifacts, including texture and background effects, show a relatively weak influence on the accuracy of the CNN. Because both of these artifacts cause changes in relative peak intensities, the distribution of misclassifications suggest that peak intensities have a more subtle role in the identification of stoichiometric single phases.

To assess the reliability of predictions made by our model, we examined the probability distributions given by the ensemble CNN. In **Fig. 3c**, we compare the probabilities of correct and incorrect classifications made when the CNN is applied to simulated spectra containing mixed artifacts. All correct classifications are accompanied by a probability greater than 70%, with an average of 93%, whereas incorrect classifications show a wide range of probabilities with a much lower average of 46%. This dichotomy suggests that probabilities are akin to confidence in the prediction and may be used as a reliable metric to gauge the likelihood that a classification is correct. If, for example, predictions are constrained to those with a probability above 70% (which comprise 84% of all spectra in the test set), then the accuracy increases from 92% to 96%. On the other hand, when the probability is lower than 70%, we propose that the model should raise a “red flag,” signifying that manual intervention is needed to clarify the identity of the underlying phase. Interestingly, even when an incorrect classification is made regarding the most probable phase, the correct phase is present within the top three suspected phases for 99% of all test spectra. Therefore, though manual intervention may occasionally be required to handle complex spectra, the problem is greatly simplified by allowing the user to choose from a small set of probable phases.

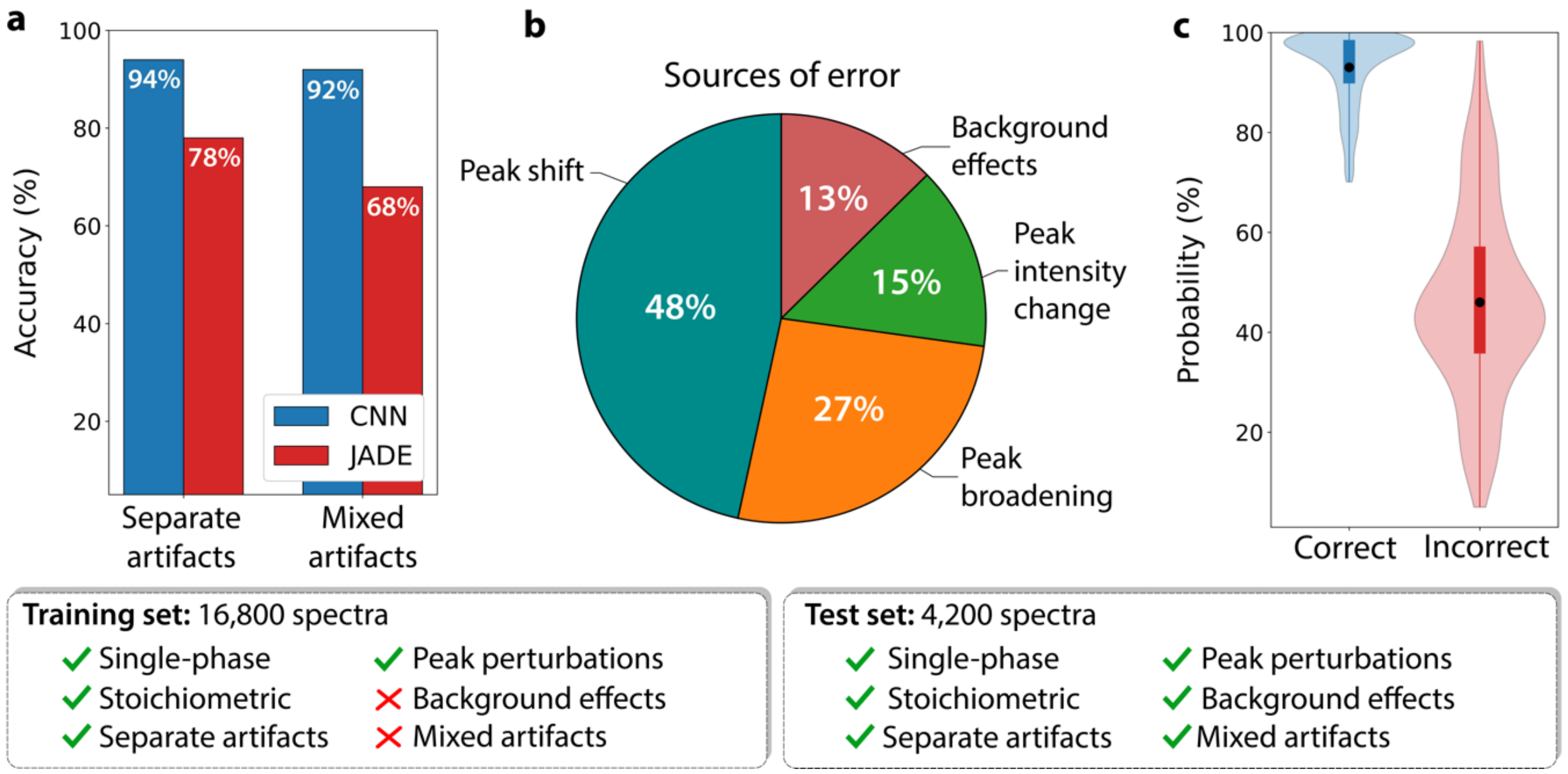

**Figure 3: (a)** The accuracies given by the CNN and JADE when applied to simulated spectra containing (i) individual artifacts applied separately and (ii) mixed artifacts applied altogether. **(b)** Sources of error in the CNN are illustrated by calculating the fraction of misclassifications that occur for spectra containing each separate artifact. **(c)** Distributions of probabilities given by the CNN when correct and incorrect classification are made during testing on spectra containing mixed artifacts. Violins plots illustrate the density of probabilities, whereas embedded boxes extend from the lower to upper quartiles. Black dots are used to denote the average probability in each case.

### Incorporating non-stoichiometry

To determine whether the accuracy of our model extends to non-stoichiometric materials, we built a test set of XRD spectra simulated from 20 experimentally reported solid solutions in the Li-Mn-Ti-O-F chemical space. These materials, listed in **Supplementary Table S3**, were manually selected from the ICSD to ensure that their compositions are different (greater than 0.05 mole fraction) than those of the stoichiometric phases already considered in the previous section. To isolate the effects of non-stoichiometry, diffraction spectra were simulated without including any experimental artifacts. We first restricted the training set to include only diffraction spectra derived from stoichiometric materials to illustrate the necessity of including additional reference phases with non-stoichiometry (i.e., from hypothetical solid solutions). Similarly, JADE was applied to the new test set containing solid solutions while restricting its reference database to contain only stoichiometric phases. In doing so, neither method can be used to predict the exact compositions of the solid solutions. Instead, their prediction accuracy can be resolved into two components: (i) Is the predicted structure isomorphic to the true structure? (ii) How similar are the predicted and true compositions? Isomorphism was verified using the pymatgen structure matcher[16]. Differences in compositions were quantified using the mole fraction distance between the barycentric coordinates of each phase in the Li-Mn-Ti-O-F chemical space (i.e., with each constituent element representing a vertex). For example, the compositional difference between $LiMnO_2$ and $LiMn_{0.5}Ti_{0.5}O_2$ is quantified as 0.125 mole fraction since 0.5 out of 4 elements are interchanged in the formula unit.

In **Fig. 4a**, we show the fraction of non-stoichiometric materials with structures correctly identified by the CNN and JADE when only stoichiometric reference spectra are used for training or profile matching. This case is labeled "Without NS" where NS denotes non-stoichiometry. The CNN correctly classifies the structures of 11/20 spectra, whereas JADE gives only 7/20 correct structural classifications. For the same set of spectra, we illustrate the differences between true compositions and those predicted by the CNN in **Fig. 4b**. Errors in the predicted compositions range from 0.05 to 0.82 mole fraction, with an average value of 0.38. Therefore, when only stoichiometric reference phases are used, neither the deep learning algorithm nor conventional profile matching techniques can be utilized to reliably predict the structure or composition of non-stoichiometric materials from their diffraction spectra. This conclusion supports our initial expectations given that substantial off-stoichiometry is known to cause large changes in the positions and intensities of diffraction peaks. Although data augmentation is useful (and necessary) to account for relatively weak deviations from ideality, it is not capable of extrapolating to larger changes well beyond those included in the training set.

A proper treatment of non-stoichiometry necessitates additional reference phases with compositions that more closely match experimentally observed solid solutions. To this end, we introduced XRD spectra simulated from hypothetical solid solutions spanning the Li-Mn-Ti-O-F space into the training set. In addition to the 21,000 spectra obtained from the 140 stoichiometric materials, 17,250 new spectra were derived from 115 hypothetical solid solutions (115 materials × 150 augmentations). Perturbations were applied *via* the data augmentation procedure described in the **Methods**, and 80% of the resulting diffraction spectra were used to re-train the CNN. For comparison, the same set of hypothetical solid solutions were also added to the reference database used by JADE. Both updated models were then applied to the test set containing 20 diffraction spectra simulated from the experimentally reported non-stoichiometric materials. The fraction of structures correctly identified by each method is displayed in **Fig. 4a**, labeled "With NS". In contrast to earlier results, the CNN and JADE achieve much higher accuracies of 95% and 70%, respectively. These improvements in performance are realized without sacrificing much accuracy in the classification of stoichiometric materials – our updated model correctly identifies 89% of phases across the test set containing simulated diffraction spectra with mixed artifacts, a decrease of only 3% compared to the CNN trained only on stoichiometric phases (**Fig. 3a**). In **Fig. 4b**, we present the updated distribution of errors in compositions given by the CNN trained with non-stoichiometric phases. Differences between the predicted and true compositions now range from 0.02 to 0.54 mole fraction, with an average value of 0.18. Hence, these results highlight the advantages of including non-stoichiometric reference phases, which nearly doubles the number of correctly identified structures and reduces compositional errors by ~50% when classifying experimentally reported solid solutions.

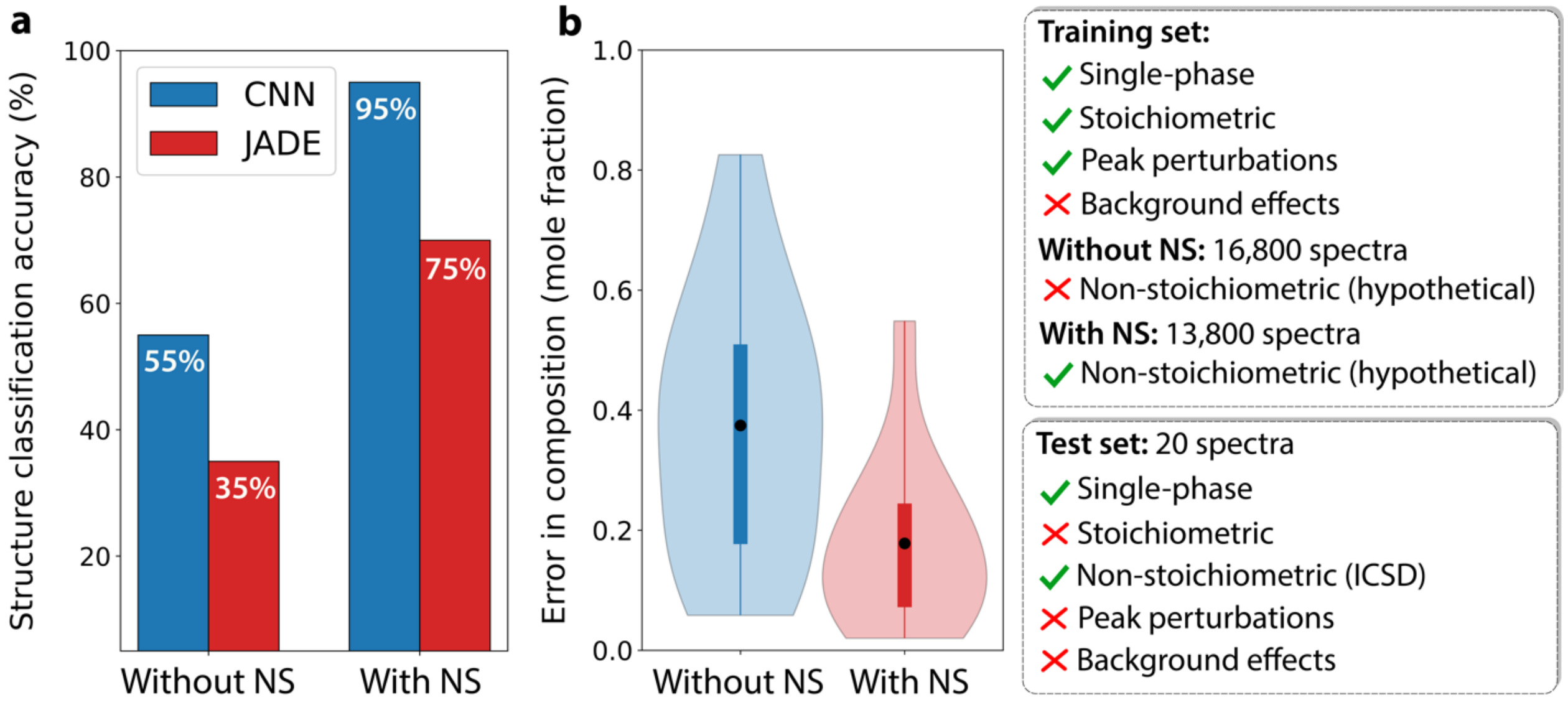

**Figure 4: (a)** For a set of diffraction spectra derived from 20 experimentally reported solid solutions, the fractions of structures correctly identified by the CNN and JADE are shown in two cases: (i) when the training set includes only stoichiometric reference phases (Without NS), and (ii) when the training set is augmented with hypothetical solid solutions (With NS). **(b)** For the same set of spectra, differences between true compositions and those predicted by the CNN are quantified by their mole fraction difference. Violin plots illustrate the full distribution of errors, whereas embedded boxes range from lower to upper quartiles. Black dots are used to denote the average probability given in each case.

## Multi-phase classification

Extending the CNN to characterize mixtures of materials, we constructed three new test sets, each containing 1,000 simulated multi-phase diffraction spectra. These tests were designed to mimic samples with multiple phases by creating linear combinations of single-phase diffraction peaks derived from 140 stoichiometric reference phases in the Li-Mn-Ti-O-F chemical space. The first two sets consider mixtures generated from randomly selected two- and three-phase combinations with equal weight fractions of the reference phases. In the last set, we probe the effects of impurity phases by simulating two-phase spectra where the weight fractions of the majority and minority phases are randomly set to constitute 70-90% and 10-30% of the mixture, respectively. In all three test cases, data augmentation is applied using mixed artifacts (peak shifting, broadening, and texture as well as a diffuse and noisy background signal) so that the resulting spectra provide an realistic representation of experimental measurements.

In addition to our newly developed branching algorithm (denoted B-CNN hereafter), multi-phase identification was performed using three other techniques for comparison: (i) based on the work of Maffettone *et al.*[12], a "single-shot" approach (S-CNN) was employed such that the two or three materials with the highest probabilities are chosen for each two- or three-phase mixture, respectively; (ii) by training the CNN explicitly on simulated multi-phase spectra (M-CNN) as described in the work of Lee *et al.*[8], entire mixtures of phases are directly predicted as opposed to separately identifying individual phases; (iii) using JADE to obtain a list of suspected phases for each mixture based on profile matching, the two or three highest-ranked materials are chosen for two- and three-phase spectra, respectively. Given that method (ii) requires many possible linear combinations of single-phase spectra to produce a sufficient number of multi-phase spectra for training, only ideal diffraction spectra were used without applying any data augmentation. Further details regarding this technique are supplied in **Supplementary Note 6**.

In **Fig. 5a**, we show the fraction of phases correctly identified by each of the four methods when tested on two- and three-phase mixtures with equally distributed weight fractions. Among all of the techniques considered here, our newly developed B-CNN algorithm achieves by far the highest accuracy, correctly identifying 87% and 78% of all materials from two- and three-phase spectra, respectively. This outperforms previously reported methods based on deep learning, with the S-CNN[12] and M-CNN[8] giving accuracies of 70% (54%) and 65% (58%) in the classification of two-phase (three-phase) mixtures. Despite their similarity in performance, these two approaches highlight separate limitations. Recall that the M-CNN does not utilize data augmentation to expand the diversity of its training set, and therefore often fails when applied to diffraction spectra containing large perturbations arising from experimental artifacts. In contrast, the S-CNN accounts for possible artifacts through physics-informed augmentation (as in our approach) and consequently is more robust against changes in the diffraction spectra. However, since the S-CNN identifies all phases in a "single shot" without subtracting known diffraction peaks, it leads to misclassifications when similar reference phases produce comparable probabilities for a given spectrum. The B-CNN improves upon both shortcomings using an iterative process of single-phase identification and profile subtraction to achieve higher accuracy. Furthermore, by maximizing the probability over all phases in the predicted mixture, the B-CNN ensures that the first iteration of phase identification is not over-prioritized. If only the most probable phase is evaluated at each step without maximizing probability over the entire mixture, lower accuracies of 78% and 69% are given across two- and three-phase mixtures, respectively.

In **Fig. 5b**, we compare the accuracy of each approach for the classification of majority/minority two-phase mixtures. The B-CNN again outperforms all other evaluated approaches. However, the reliability of our model varies substantially in the identification of majority versus minority phases. The B-CNN correctly classifies 92% of all majority phases, matching its performance across single-phase spectra and therefore suggesting the presence of impurity phases has little to no effect on majority phase identification. Identifying minority phases, on the other hand, presents a greater challenge, as reflected by a lower accuracy of 64% given by the B-CNN. We note that most misclassifications occur due to imperfect applications of profile subtraction that occasionally leave behind residual intensities or subtract some diffraction peaks associated with the minority phase of interest. Despite this limitation in the *identification* of minority phases, the model generally performs reliably in their *detection*. Recall that the number of phases in a mixture is determined by halting the B-CNN when all diffraction intensities fall below 5% of the initially measured maximum intensity. With this cutoff, the B-CNN correctly reports the presence of a second phase in 93% of the two-phase mixtures with unequally distributed weight fractions. For comparison, when the B-CNN is applied to simulated single-phase spectra with mixed artifacts (**Fig. 3a**) using the same cutoff intensity of 5%, the number of phases is overestimated in only 9% of the samples. The key component enabling a reliable prediction for the number of phases is the approach to profile subtraction. Here, known diffraction peaks are fit to the spectrum through DTW so that their subtraction yields a new spectrum that accurately represents the mixture minus the phase(s) that has already been identified. This capability is particularly useful in the optimization of synthesis procedures, where it is of interest to know whether the formation of a targeted product is accompanied by some impurity phase.

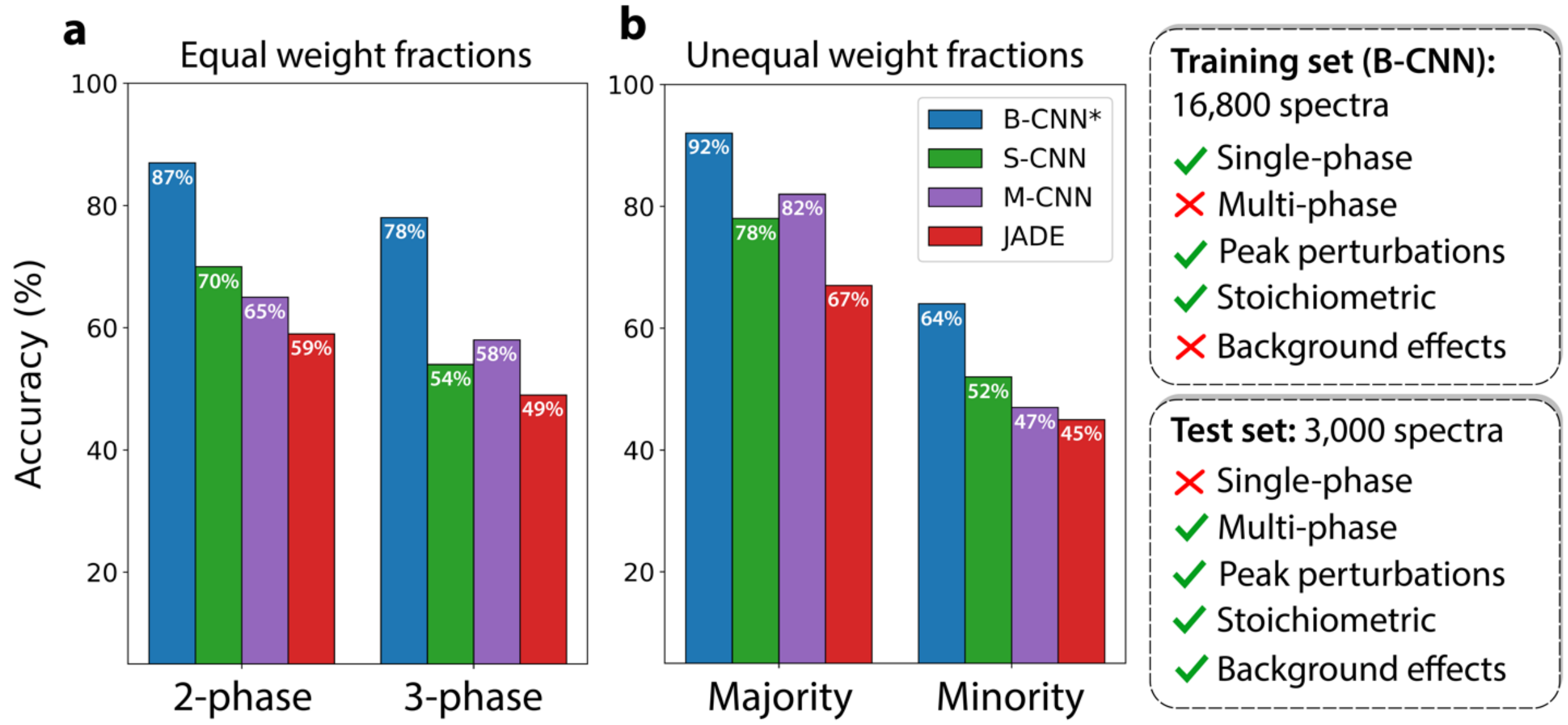

**Figure 5: (a)** The fractions of phases correctly identified by the B-CNN (*introduced in this work) when applied to simulated diffraction spectra of two- and three-phase mixtures with equally distributed weight fractions. For comparison, accuracies obtained using two methods based on previous work (S-CNN[12] and M-CNN[8]) are shown, in addition to results from JADE. **(b)** These same techniques are applied to diffraction spectra of two-phase mixtures with unequally distributed weight fractions of 10-30% and 70-90%. Accuracies are divided into the identification of majority and minority phases.

**Application to experimental spectra**

As a final demonstration of the generalizability of our approach, the B-CNN was applied to experimentally measured spectra in the Li-Mn-Ti-O-F chemical space. In **Table 1**, we list the fraction of phases correctly identified by the CNN versus JADE, with results categorized by the artifacts and number of phases included for each class of spectra (previously described in **Experimental measurements**). For the classification of pristine diffraction spectra, the CNN correctly identifies all ten phases considered. Interestingly, JADE incorrectly classifies one material ($Li_2TiO_3$) from this category. Upon further inspection, the error is attributed to large deviations in the relative peak intensities between the measured and ideal spectra of $Li_2TiO_3$ (shown in **Fig. S2**), possibly caused by stacking faults in the sample[23]. In the analysis of spectra with diffuse and noisy background signals, the CNN correctly identifies all but one material (anatase $TiO_2$), likely due to the fact that it exhibits significant diffraction peaks at low values of $2\theta$ where the amorphous background is strong. JADE is found to be more sensitive to background effects as it yields five misclassifications across these 20 spectra. These misclassifications occur because JADE fails to index peaks that blend in with the background signal and have low intensities or broad widths after a baseline correction is applied. The CNN is more robust against these perturbations since it is trained on spectra having diffraction peaks with varied intensities and widths.

For spectra containing peak shifts, the CNN correctly identifies five out of six phases. In contrast, JADE struggles to handle changes in peak positions, identifying only two phases from this category. This highlights a key weakness of profile matching techniques, which fail when there is weak overlap between measured and simulated diffraction peaks owing to a shift in $2\theta$. Fortunately, because the CNN can handle these changes through data augmentation, its performance remains reliable in the classification of spectra with peak shifts. When diffraction peaks are broadened, the CNN and JADE correctly identify five and four phases, respectively, from the five measured spectra. The single misclassification from JADE occurs for $Li_2MnO_3$ owing to a strong overlapping of its neighboring diffraction peaks, an effect which is accounted for by the CNN during training. For the six spectra with changes in their peak intensities, the CNN correctly classifies five phases while JADE identifies four. The misclassification made by the CNN occurs because the varied peak intensities closely resemble those of a hypothetical solid solution ($Li_{0.5}Mn_{1.5}TiO_4$) that is isomorphic to the true phase ($LiMnTiO_4$). Across non-stoichiometric materials, the CNN correctly predicts all four materials to adopt the rocksalt structure, whereas JADE finds only three phases to be rocksalt. For both methods, the predictions are facilitated by the introduction of hypothetical solids solutions; without including these additional reference phases, neither the CNN nor JADE predicts any of the four samples to be rocksalt-structured.

For the classification of multi-phase mixtures, JADE provides limited accuracy. Only 7/10 and 9/15 phases are correctly identified from two- and three-phase spectra, respectively. Such limitations in accuracy can be attributed to the inability of profile matching techniques to distinguish between diffraction peaks produced by several phases, which often overlap with one another. The B-CNN adeptly overcomes these limitations and correctly identifies 10/10 and 13/15 phases in the two- and three-phase mixtures, respectively. Hence, the benefits provided by deep learning are highlighted by the noticeable disparity between the performance of the CNN versus

JADE, especially when applied to multi-phase spectra. This advantage is vital to assist in targeted synthesis, considering that attempts to produce novel inorganic materials are frequently impeded by the appearance of multiple impurity phases. Our deep learning approach can therefore be used to identify not only desired products, but also impurity phases, which provide insight into why a given synthesis procedure failed and inform future attempts.

The results from testing the CNN on experimentally measured spectra (**Table 1**) closely match the performance on simulated spectra (**Figs. 3-5**). For example, in spectra where we include a single type of artifact, the CNN correctly identifies 94% of phases from both simulated and experimentally measured single-phase spectra. This lends credence to the simulation-based test cases that are rich in data (e.g., a total of 4,200 single-phase test spectra were derived from stoichiometric materials) and suggests that the simulated spectra used for training and testing provide a realistic representation of experimental measurements.

**Table 1:** Fractions of materials correctly identified by the CNN and JADE when applied to experimentally measured XRD spectra designed to sample possible artifacts arising during sample preparation and synthesis. For diffraction spectra of non-stoichiometric materials, a classification is considered correct if the predicted structure is isomorphic to the true structure.

| Experimental procedure | Anticipated artifact | CNN | JADE |
|---|---|---|---|
| Single-phase | | | |
| Pristine samples | None | 10/10 | 9/10 |
| Kapton tape overlaid | Diffuse baseline | 9/10 | 8/10 |
| Rapid XRD scan | Noisy baseline | 10/10 | 7/10 |
| Thick samples | Shifts in 2θ | 5/6 | 2/6 |
| Ball milled | Broadening | 5/5 | 4/5 |
| Partially disordered | Intensity variation | 5/6 | 4/6 |
| Solid solutions | Non-stoichiometry | 4/4 | 3/4 |
| Multi-phase | | | |
| Two-phase mixtures | None | 10/10 | 7/10 |
| Three-phase mixtures | None | 13/15 | 9/15 |
| | **Overall accuracy:** | **71/76 (93.4%)** | **53/76 (71.4%)** |

## Discussion

In summary, we developed an improved deep learning technique that can reliably automate the identification of inorganic materials from XRD spectra. A key advantage of our approach is the physics-informed data augmentation procedure that accounts for several experimental artifacts commonly observed after sample preparation and synthesis. Conventional profile matching techniques often fail when materials variations cause large differences between observed and simulated diffraction peaks, requiring manual intervention to analyze any irregularities and identify the samples of interest. In contrast, our CNN learns these differences during training, and therefore can autonomously perform phase identification from complex spectra. These benefits are highlighted by the test results presented in this work, which show that the performance of profile matching quickly deteriorates as larger perturbations are applied to the diffraction spectra, whereas the CNN remains reliable in the presence of such perturbations. Furthermore, even though our model is trained only on spectra that account for three types of artifacts (strain, texture, and domain size), it is demonstrated to successfully generalize to spectra outside of the training set. For example, our algorithm achieves a high accuracy for the identification of spectra with diffuse and noisy baseline signals, as well as for samples containing unexpected artifacts (e.g., possible stacking faults in $Li_2TiO_3$).

Of the artifacts considered in our work, changes in peak positions are shown to be the most challenging to deal with, comprising nearly half of all misclassifications made by the CNN when applied to the simulated diffraction spectra of single-phase stoichiometric materials. Because peak positions are derived from the spacings between crystallographic planes, and therefore the lattice parameters of the material, it is difficult to distinguish between isomorphic phases when their structures have a significant degree of strain. We find that our model provides an optimal treatment of changes in peak positions by including samples with as much as $\pm 4\%$ strain in the training set, which is unlikely to be exceeded in experiment unless the materials contain substantial off-stoichiometry. Indeed, tests involving an increased magnitude of strain in the training set led to decreased accuracy during testing owing to degeneracies between the diffraction spectra of similar phases. In general, the bounds used for data augmentation should reflect the experimental system at hand; for example, larger perturbations may be beneficial in cases where certain artifacts are expected to dominate (e.g., epitaxial strain in thin films). To avoid degeneracy of spectra in the training set, the number of reference phases should be constrained to include only those that are expected to arise in experiment – for synthesis, these can be chosen to reflect the composition space spanned by the precursors used and the possibility of reactions with oxygen, water, or $CO_2$ in air.

The importance of peak positions is further highlighted by our tests involving non-stoichiometric materials. Varying the composition of a material typically leads to changes in its lattice parameters, which in turn shifts the positions of its diffraction peaks. As a result, when the CNN is trained only with stoichiometric reference phases, it frequently fails to identify the structures of non-stoichiometric materials. Because the model is trained to identify individual phases, rather than their symmetry, it does not necessarily learn the subtle relationships between peak positions imposed by the space group of each structure. Instead, it considers the positions of all peaks and makes a comparison with known phases in the training set. Therefore, when non-stoichiometry causes large shifts in the positions of diffraction peaks, the CNN will struggle if it has no reference phase available with comparable peak positions. With this in mind, we improved the treatment of non-stoichiometric materials by building a library of hypothetical solid solutions following Vegard's law. After adding their diffraction spectra to the training set, the CNN correctly identifies the structures for 95% of the non-stoichiometric materials considered during testing. We note that this approach is successful because the lattice parameters of most solid solutions follow Vegard's law with only minor deviations[24]. When deviations do occur, data augmentation ensures that the match between hypothetical and experimentally observed phases need not be exact for the model to maintain a high level of accuracy for the identification of the material's structure.

Despite the improved prediction of structure enabled by introducing hypothetical solid solutions to the training set, predicting the compositions of non-stoichiometric materials remains challenging. This limitation can be understood by considering the effects of non-stoichiometry on diffraction peak intensities, which are influenced by the structure's internal cell coordinates and site occupancies. Given the similarity of structural frameworks between materials forming solid solutions, changes in cell coordinates are usually small and therefore do not contribute significantly to differences in peak intensities. Changes in site occupancies, however, strongly influence peak intensities owing to the distinct scattering factors of substituted species. As opposed to changes in lattice parameters that can be described by Vegard's law, an automatic prediction of site occupancy is more difficult to achieve because site occupancies can redistribute in solid solutions. For example, partial inversion (i.e., swapping Wyckoff positions) between lithium and transition metal ions has been observed in spinel $LiMn_{2-x}Ti_xO_4$[25]. Such differences give rise to errors in predicted compositions, not structures, because site occupancies control peak intensities while leaving peak positions relatively unaffected. Hence, we reiterate that our approach is not designed to give precise refinements of composition, but rather to provide a reliable prediction of structure and an estimate of composition. Beyond the scope of this work, future efforts may be conducted to design a more accurate prediction of site occupancies so that refinement can be carried out autonomously. A recent report by Mattei *et al.* has shown some progress toward this end, providing an approach to enumerate many possible distributions of site occupancies with the goal of identifying the best match with experimental measurements[26]. As their approach requires that the structural framework of the suspected phase be known prior to refinement, our model may prove useful in coordination with their algorithm.

When samples contain more than one material, new challenges arise as diffraction peaks often overlap and can be difficult to distinguish. To handle multi-phase spectra, we designed a branching algorithm that iterates between phase identification and profile subtraction to identify the combination of phases that maximizes the average probability given by the CNN. This approach yields exceptionally high accuracy across simulated and experimentally measured multi-phase XRD spectra, exceeding the performance of profile matching techniques and recently published methods based on deep learning. The advantages of our branching algorithm can be summarized by two main points. First, by training only on single-phase spectra, we avoid the combinatorial explosion of training samples that would arise if multi-phase spectra were instead used. Because the number of pristine reference spectra is kept low, many experimental artifacts can be included through physics-informed data augmentation, which ensures the model is robust against perturbations in diffraction spectra caused by defects or impurities. Second, our algorithm avoids confusion between phases with similar reference spectra by identifying phases in a one-by-one manner and iteratively subtracting their diffraction peaks from the spectrum until all non-negligible intensities have been accounted for. The removal of known peaks prevents the algorithm from overestimating the number of phases in a sample, which would otherwise occur if the probability distribution given by the CNN was assumed to represent a mixture of phases (e.g., assuming all phases with a probability $\geq 50\%$ exist in a given sample).

## Conclusion

We have demonstrated that a deep learning algorithm based on a CNN can be trained to identify inorganic materials from complex diffraction spectra. Physics-informed data augmentation was shown to accurately account possible experimental artifacts in measured diffraction spectra, therefore improving the generalizability of the CNN. Simulated spectra derived from hypothetical solid solutions were also added to the training set, which improves the performance of the model when dealing with off-stoichiometric samples. For samples containing multiple phases, an iterative process of phase identification and profile subtraction was designed to maximize the probability given by the CNN over all phases in the predicted mixture, which performs well when applied to multi-phase spectra. The proposed accuracy of our deep learning approach was validated with respect to simulated and experimentally measured diffraction spectra.

Although our current tests focus on materials in the Li-Mn-Ti-O-F space, the algorithm developed here (provided below in **Code Availability**) can be applied to any arbitrary composition space given a set of reference phases, which can be extracted from existing crystallographic databases. Because the number of training samples required by our method scales linearly with the number of reference phases, and only 150 spectra are generated for each phase, the entire process of spectrum simulation and CNN training can be extended to broad composition spaces without requiring excessive resource use. For example, based on the 140 reference phases in the Li-Mn-Ti-O-F space, a completely new model can be built from scratch in about one day using 16 CPUs. Therefore, given the efficiency of our approach and the promising results illustrated throughout this work, we suggest that the algorithm developed here may be used to effectively accelerate materials discovery by incorporating automatic phase identification to support high-throughput and autonomous experimental workflows.

## Code availability

A public repository containing the methods discussed in this work can be found at https://github.com/njszym/XRD-AutoAnalyzer. This includes the code used to perform data augmentation, generation of hypothetical solid solutions, training of the CNN, and application of the CNN to classify XRD spectra using the probabilistic branching algorithm. A pre-trained model is available for the Li-Mn-Ti-O-F chemical space.

## Data availability

All XRD spectra used for testing can be found on Figshare. Reported accuracies can be reproduced by applying our pre-trained model to these spectra.

## Acknowledgements

This work was supported as part of the Joint Center for Energy Storage Research (JCESR), an Energy Innovation Hub funded by the U.S. Department of Energy, Office of Science, Basic Energy Sciences.

**Supplementary Information**

# A probabilistic deep learning approach to automate the interpretation of multi-phase diffraction spectra

Nathan J. Szymanski[1,2], Christopher J. Bartel[1,2], Yan Zeng[2], Qingsong Tu[2], and Gerbrand Ceder[1,2*]

**Information included:**

Supplementary Tables S1-S3

Supplementary Figures S1-S2

Supplementary Notes 1-6

[1]Department of Materials Science & Engineering, UC Berkeley, Berkeley, CA 94720, USA

[2]Materials Sciences Division, Lawrence Berkeley National Laboratory, Berkeley, CA 94720, USA

*correspondence to gceder@berkeley.edu

**Supplementary Table S1:** Stoichiometric reference phases from the Li-Mn-Ti-O-F composition space that are reported in the ICSD.

| Formula | Space group no. | ICSD ID |
|---|---|---|
| Li2TiF6 | 136 | 256029 |
| Ti4O7 | 2 | 6098 |
| LiTiMnF6 | 150 | 69047 |
| Mn | 141 | 163245 |
| TiOF5 | 15 | 32676 |
| Ti6O11 | 2 | 9039 |
| LiMnF4 | 14 | 62655 |
| Li3Ti4O8 | 3 | 151917 |
| MnO | 225 | 643192 |
| MnF2 | 136 | 68736 |
| MnO2 | 164 | 53991 |
| Ti | 229 | 76165 |
| Ti7O13 | 2 | 9040 |
| MnO | 186 | 262928 |
| MnO3 | 74 | 173645 |
| Li2MnF5 | 15 | 202394 |
| TiO2 | 205 | 189326 |
| Mn | 225 | 675395 |
| LiMnO2 | 59 | 84642 |
| Ti | 72 | 672745 |
| LiTi2O4 | 58 | 182579 |
| TiO2 | 189 | 41056 |
| Mn | 191 | 673020 |
| TiO2 | 35 | 97008 |
| Li3TiF6 | 15 | 405346 |
| TiO | 225 | 670890 |
| Li2O2 | 129 | 26892 |
| LiMn2O4 | 70 | 54701 |
| Li4TiO4 | 63 | 75164 |
| Mn2O5 | 55 | 63642 |
| LiF | 225 | 674429 |
| Ti3O | 149 | 36055 |
| Li | 229 | 674846 |
| MnF3 | 15 | 73113 |
| Li2MnO3 | 15 | 21022 |
| TiO2 | 139 | 671268 |
| Ti3O | 162 | 23575 |
| LiMnO4 | 63 | 89505 |
| Ti3O | 163 | 24082 |
| TiMnO3 | 148 | 184649 |
| Ti | 139 | 671047 |
| MnO2 | 84 | 163598 |
| Mn3O4 | 74 | 672421 |
| Li2Ti6O13 | 12 | 182966 |
| LiTi2O4 | 12 | 180011 |
| Li2TiO3 | 15 | 257005 |
| MnF2 | 225 | 672195 |
| Li | 213 | 161377 |
| Li2Ti3O7 | 11 | 193803 |
| TiMnO3 | 161 | 184650 |
| TiMnO3 | 62 | 158732 |
| Mn5O8 | 12 | 16956 |
| Ti4O5 | 87 | 174033 |
| MnO2 | 87 | 20227 |
| Ti2O | 164 | 99784 |
| TiF2 | 225 | 68400 |
| LiTi2O4 | 11 | 182580 |
| Li | 166 | 426951 |
| Li2O | 225 | 671967 |
| TiF3 | 221 | 28783 |
| Li2MnO3 | 12 | 187500 |
| LiTi2O4 | 227 | 154982 |
| Li | 194 | 671115 |
| Ti6O | 162 | 20042 |
| LiMnO2 | 141 | 40486 |
| Mn3O4 | 141 | 77478 |
| Ti6O | 163 | 17009 |
| MnF2 | 62 | 672705 |
| Li | 220 | 109012 |
| Ti2O3 | 167 | 9646 |
| TiMn2 | 194 | 197772 |
| TiF3 | 167 | 52164 |
| Ti3O5 | 12 | 647543 |
| Ti3Mn3O | 227 | 29052 |
| MnF2 | 60 | 672197 |
| LiTiO2 | 141 | 164158 |
| MnF2 | 61 | 672706 |
| Mn2O7 | 14 | 60821 |
| LiO2 | 58 | 180561 |
| MnO2 | 62 | 171866 |
| Mn2O3 | 148 | 236254 |
| Ti3O5 | 13 | 194465 |
| TiO2 | 62 | 671269 |
| Li2MnF6 | 164 | 15791 |
| LiTiO2 | 227 | 48128 |
| Ti2O3 | 62 | 187466 |
| Mn2O3 | 206 | 290637 |
| Li2O2 | 14 | 671295 |
| Ti3O7 | 12 | 5455 |
| MnF2 | 205 | 672196 |
| TiO | 12 | 56694 |
| Li2Ti(OF3)2 | 12 | 2558 |
| MnO2 | 136 | 670366 |
| Mn2O3 | 61 | 290641 |
| LiMn4O8 | 198 | 252035 |
| TiO2 | 61 | 77693 |
| TiO2 | 136 | 168138 |
| Li2O2 | 194 | 180557 |
| LiTi2O4 | 31 | 182581 |
| Mn3O4 | 57 | 188903 |
| Li | 225 | 76948 |
| MnO2 | 58 | 27789 |
| Mn2O3 | 199 | 33647 |
| Li2O | 166 | 108886 |
| Ti3O5 | 15 | 35148 |
| TiO2 | 60 | 189320 |
| LiMn2O4 | 227 | 192369 |
| Ti5O9 | 1 | 653560 |
| Mn | 216 | 187036 |
| TiO | 189 | 196273 |
| TiMn2O4 | 95 | 22313 |
| Mn | 217 | 426954 |
| Li2MnO2 | 164 | 37327 |
| TiO2 | 12 | 41056 |
| LiO3 | 44 | 180565 |
| TiO2 | 225 | 189325 |
| MnO2 | 227 | 193445 |
| Ti6O11 | 12 | 90958 |
| Ti5O9 | 2 | 31401 |
| Mn | 229 | 5392 |
| MnO2 | 14 | 40486 |
| LiO2 | 12 | 159510 |
| TiO2 | 152 | 24073 |
| Ti | 194 | 672692 |
| MnF4 | 88 | 62068 |
| Mn2O2F9 | 15 | 26399 |
| Li | 64 | 182499 |
| Ti | 225 | 671773 |
| Ti9O17 | 2 | 9042 |
| Ti | 191 | 675189 |
| MnF2 | 111 | 12167 |
| MnO2 | 10 | 99593 |
| TiF4 | 62 | 78737 |
| Ti3O5 | 63 | 50984 |
| Mn | 213 | 163414 |
| MnO2 | 12 | 150462 |
| TiO2 | 141 | 673140 |
| Ti8O15 | 2 | 9041 |
| Ti21Mn25 | 167 | 600102 |
| Ti2MnO4 | 227 | 22383 |

**Supplementary Table S2:** Hypothetical solid solutions in the Li-Mn-Ti-O-F composition space.

| Formula | Space group no. |
|---|---|
| Ti1.5Mn0.5 | 139 |
| Li3Ti1.5Mn1.5F12 | 136 |
| Li4Ti3MnO8 | 141 |
| Li3TiOF3 | 225 |
| Ti10.5Mn1.5O18 | 148 |
| Li0.5Mn3.5F8 | 14 |
| TiMn | 139 |
| Li0.5Ti1.5 | 194 |
| Ti0.5Mn1.5 | 229 |
| Ti10.5Mn1.5O18 | 161 |
| Li2Ti9Mn(O3F)5 | 15 |
| Li2Ti3MnO8 | 227 |
| LiTi | 139 |
| Li2Ti3MnO8 | 70 |
| Li3Ti(OF)2 | 227 |
| Ti3Mn(O3F)2 | 60 |
| TiMn3(OF3)2 | 60 |
| LiTi3 | 122 |
| TiMn3(OF3)2 | 205 |
| LiTiMn6O8 | 141 |
| Li3Ti(OF)2 | 141 |
| Li2Ti9Mn(O3F)5 | 13 |
| TiMn3F12 | 15 |
| Li6Ti3Mn3(OF3)5 | 15 |
| LiTi3Mn2O8 | 95 |
| TiMn(OF)2 | 136 |
| Li3MnOF3 | 225 |
| LiTi3 | 225 |
| Ti1.5Mn0.5O4 | 136 |
| LiMn3F8 | 14 |
| Li5Ti3(O3F)2 | 141 |
| Ti3Mn | 72 |
| Ti0.5Mn1.5O4 | 58 |
| LiTi8Mn3O16 | 227 |
| Ti3Mn(O3F)2 | 205 |
| LiTi3O3F | 225 |
| Ti7.5Mn4.5O18 | 161 |
| Ti7.5Mn4.5O18 | 148 |
| Ti3Mn3O8 | 227 |
| Ti3MnO4 | 225 |
| TiMn5O8 | 141 |
| Li3Ti | 225 |
| LiTi2Mn3O8 | 74 |
| LiTi | 229 |
| Ti3Mn | 225 |
| LiTiMn2O4 | 141 |
| LiTi2Mn9O16 | 141 |
| LiMnOF | 225 |
| Li1Ti0.5Mn4.5F12 | 136 |
| Li2TiMnO4 | 141 |
| Li3Ti3Mn2O8 | 141 |
| Ti1.5Mn0.5O3F1 | 136 |
| Li3Ti | 220 |
| Ti3MnF12 | 167 |
| Li5Ti3(O3F)2 | 227 |
| Li2TiMn3O8 | 70 |
| TiMn | 229 |
| LiTi3 | 220 |
| Ti1.5Mn0.5 | 229 |
| Li1.5Ti0.5 | 194 |
| TiMn5O8 | 74 |
| LiTiOF | 225 |
| LiTi2Mn3O8 | 141 |
| TiMnO4 | 136 |
| TiMn | 225 |
| Ti7Mn5O16 | 95 |
| Ti0.5Mn1.5O1F3 | 136 |
| Li7Ti(OF3)2 | 141 |
| TiMn3 | 225 |
| Li3Ti6Mn3O16 | 141 |
| Ti0.75Mn2.25 | 191 |
| Ti3Mn3O8 | 95 |
| TiMnO4 | 58 |
| LiTi | 225 |
| TiMn(OF)2 | 205 |
| LiMn3O3F | 225 |
| Ti3MnF12 | 221 |
| LiTi | 15 |
| LiTi5Mn6O16 | 95 |
| LiTi | 220 |
| LiTi3 | 72 |
| Li3Ti7Mn2O16 | 95 |
| LiTi | 194 |
| TiMn | 15 |
| Ti0.5Mn1.5O4 | 136 |
| Li2TiMn3F12 | 136 |
| Li1.5Ti0.5 | 229 |
| TiMn3 | 72 |
| TiMnO2 | 225 |
| Li0.5Ti1.5 | 139 |
| Li7Ti(OF3)2 | 227 |
| Li3Ti6Mn3O16 | 227 |
| TiMn3O4 | 225 |
| Li1.5Mn2.5F8 | 14 |
| LiTiMnO4 | 70 |
| TiMn3 | 15 |
| Ti1.5Mn1.5 | 191 |
| Li3Ti | 15 |
| LiTiMnO4 | 227 |
| TiMnF6 | 15 |
| Li0.5Ti1.5 | 229 |
| TiMn(OF)2 | 60 |
| TiMn | 72 |
| LiTi2Mn9O16 | 74 |
| Ti3MnO6 | 161 |
| Ti3MnO6 | 148 |
| Li3Ti8MnO16 | 227 |
| TiMn2O4 | 141 |
| Li2Ti3Mn(OF)5 | 15 |
| LiTi4MnO8 | 227 |
| Li4TiMn3O8 | 141 |
| Ti5Mn7O16 | 95 |
| Ti2.25Mn0.75 | 191 |
| Li2TiMn3O8 | 227 |
| Li2Ti3Mn(OF)5 | 13 |

**Supplementary Table S3:** A list of the 20 experimentally reported solid solutions in the Li-Mn-Ti-O-F chemical space that are taken from the ICSD and used during testing.

| Formula | Space group no. | ICSD ID |
|---|---|---|
| Li0.5Mn0.5O | 225 | 163493 |
| Li0.05Mn0.95O | 225 | 98165 |
| Mn0.15Ti0.85 | 229 | 104991 |
| Mn1.67Ti1.33 | 194 | 101107 |
| Mn3.6Ti2.4 | 194 | 104990 |
| Mn1.8Ti1.2 | 194 | 198909 |
| Li1.41Mn1.49O4 | 227 | 88138 |
| Li1.16Mn0.84O4 | 227 | 84756 |
| LiMn1.9Ti0.1O4 | 227 | 50429 |
| LiMn1.8Ti0.2O4 | 227 | 50431 |
| LiMn1.6Ti0.4O4 | 227 | 50432 |
| LiMnTiO4 | 227 | 166742 |
| Li1.06Mn1.75Ti0.28O4 | 227 | 155560 |
| Li0.94Mn1.5Ti0.5O4 | 227 | 155561 |
| Li1.33Mn1.33Ti0.33O4 | 227 | 180027 |
| Li1.23Mn0.3Ti1.47O4 | 227 | 192988 |
| Li1.3Mn0.1Ti1.6O4 | 227 | 192987 |
| LiMn0.5Ti1.5O4 | 212 | 154145 |
| LiMn0.7Ti1.3O4 | 212 | 154416 |
| LiMn0.8Ti1.2O4 | 212 | 154417 |
| LiMnTiO4 | 212 | 238492 |

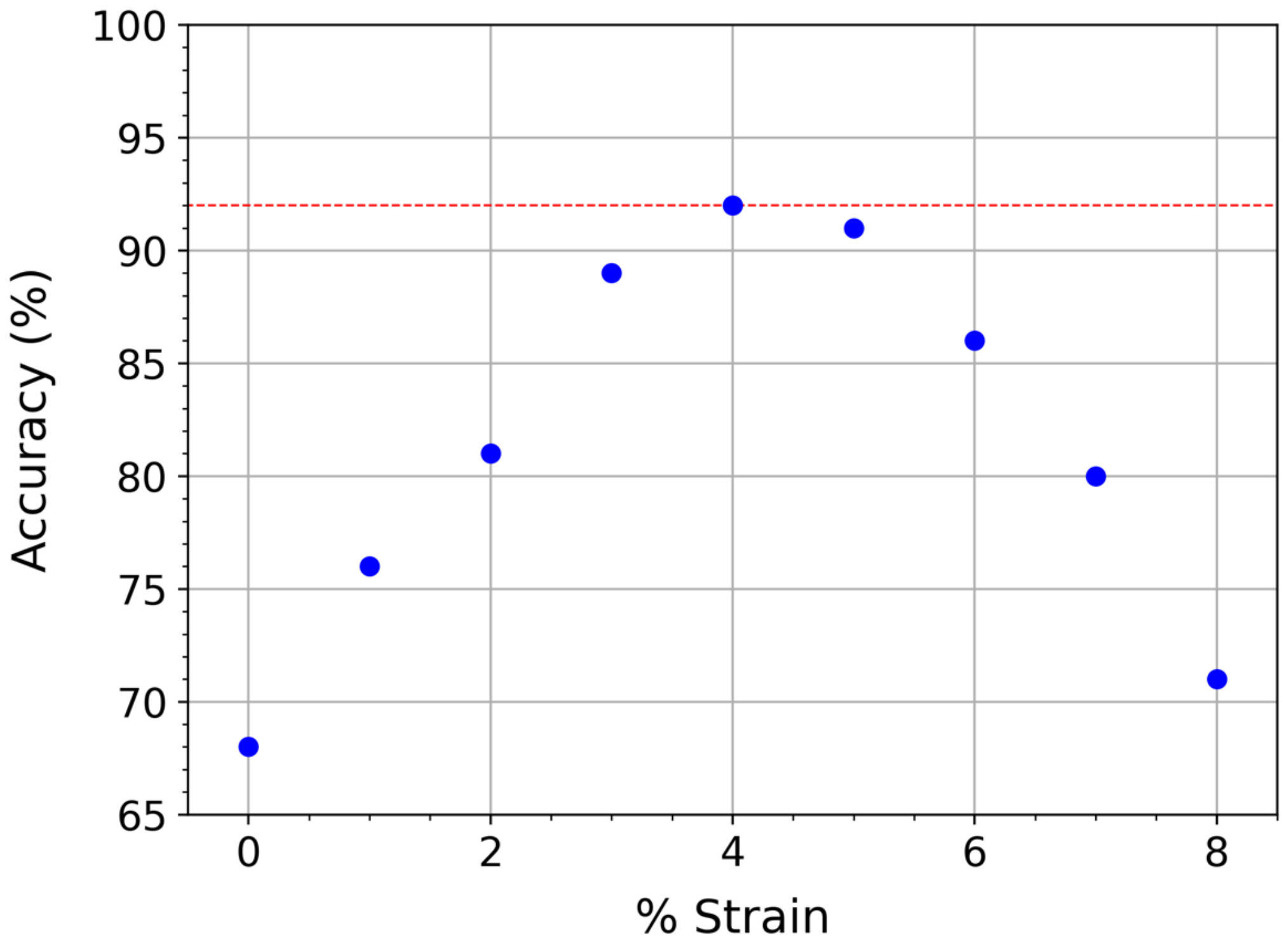

**Figure S1:** The percentage of phases correctly identified by the CNN when applied to test spectra containing strain as large as ±4%. Each blue dot represents the accuracy reported by a distinct model, which was trained on spectra derived from structures with strain as large as the value indicated by the x-axis. The red line shows the optimum accuracy (92%) that was achieved using a maximum strain of ±4% in the training set.

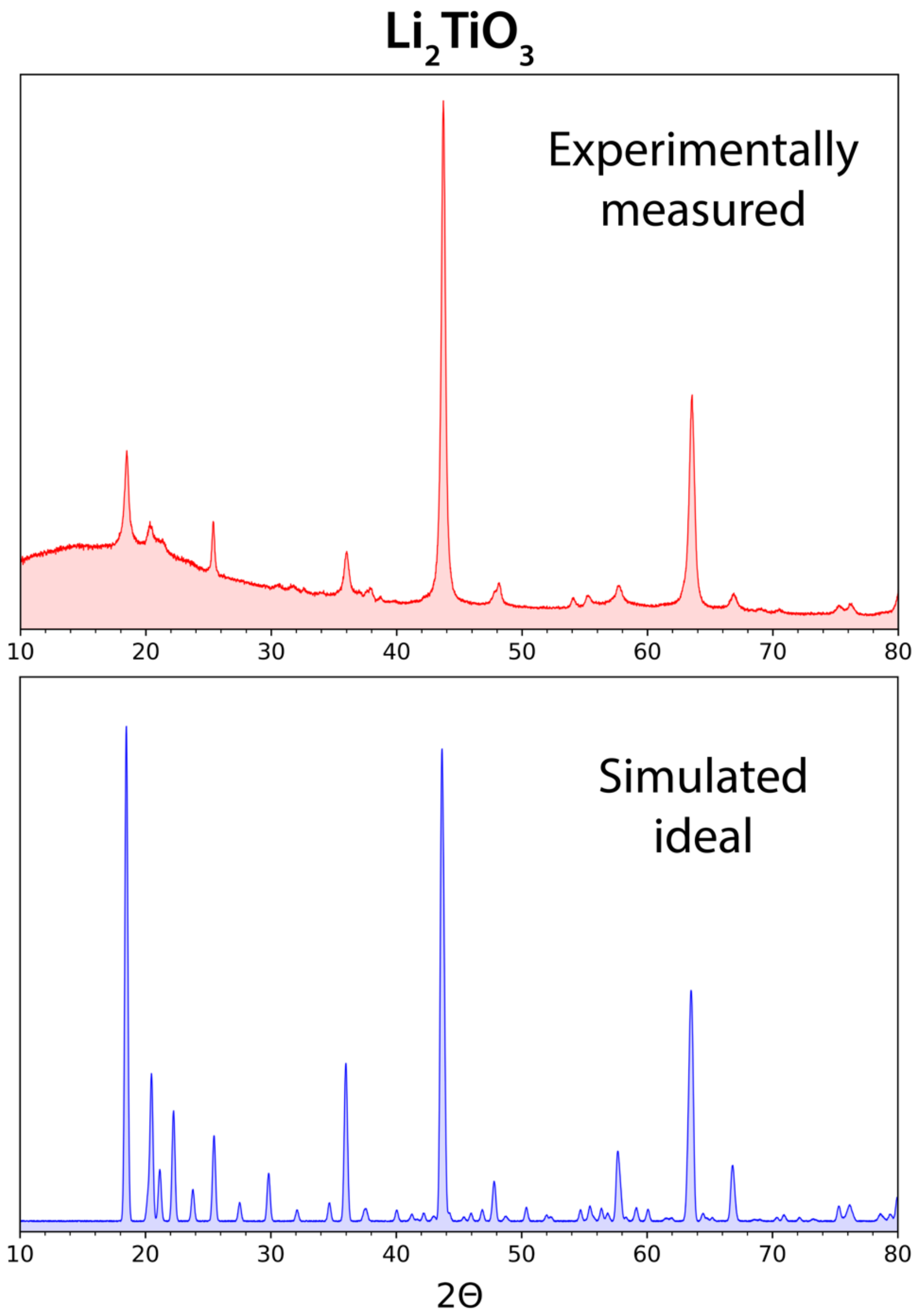

**Figure S2:** Experimentally measured (top panel) and simulated (bottom panel) spectra for $Li_2TiO_3$, showing clear differences in relative peak intensities that lead to a misclassification by JADE.

**Supplementary Note 1: Spectrum simulation and data augmentation**

**Ideal XRD spectra:** For each phase, the structure factor and Lorentz polarization factor were simulated using the XRDCalculator module from pymatgen assuming Cu $K_\alpha$ radiation. This yields a discrete list of peak positions and intensities that represent the ideal XRD spectrum. To obtain a continuous spectrum from this list, Gaussian functions were fit to the diffraction peaks such that the maximum value of each function matches the corresponding peak intensity. The full width at half maximum (FWHM) of the Gaussian was set to 0.015° to reflect narrow diffraction peaks measured from high-purity samples. The highest diffraction peak produced by any given phase was set to 100 so that all spectra display comparable intensities. Stochastic noise ranging from 0 to 1 was added to the spectrum to emulate measurements obtained experimentally.

**Data augmentation:** Three changes to the simulated spectra were considered:

1) Shifts in peak positions: Prior to calculating the XRD spectrum as described above, strain was applied to through the application of a strain tensor taking the form:

$$\begin{pmatrix} 1+\Delta c_{11} & \Delta c_{12} & \Delta c_{13} \\ \Delta c_{21} & 1+\Delta c_{22} & \Delta c_{23} \\ \Delta c_{31} & \Delta c_{32} & 1+\Delta c_{33} \end{pmatrix}$$

   Deviations from the identity matrix were obtained by randomly sampling the coefficients such that $\Delta c_{ij} \in [-0.04, 0.04]$. In all cases, the relative values of the coefficients were restricted such that the symmetry of the structure was preserved upon the application of strain. In a cubic structure, for example, the following relations must hold:

$$\Delta c_{11} = \Delta c_{22} = \Delta c_{33}$$

$$\Delta c_{ij} = 0 \text{ for } i \neq j$$

2) Varied peak intensities: To replicate texture along a preferred crystallographic plane, the indices of each diffraction peak were scaled by taking a scalar product with randomly chosen Miller indices $(hkl)$ where $h, k, l \in \{0, 1\}$. Normalization was applied such that peak intensities were scaled by as much as $\pm 50\%$ of their original values. In other words, when peak indices are completely out of phase with the preferred direction, the associated intensity is multiplied by 0.5, whereas peaks with indices completely in phase with the preferred direction have intensities multiplied by 1.5.

3) Broadening of peak widths: The FWHM ($\beta$) was modified for all peaks according to the Scherrer equation:

$$\tau = \frac{K\lambda}{\beta cos\theta}$$

The domain size ($\tau$) was randomly sampled between 1 nm and 100 nm. The form factor ($K$) was chosen to be equal to one. The wavelength $\lambda$ was set to 1.5406 Å to reflect Cu $K_\alpha$ radiation. The diffraction angle ($\theta$) is pre-defined by each peak position.

## Supplementary Note 2: Convolutional neural network architecture and training

**Architecture:** The input layer of the CNN is one-dimensional and contains 4,501 values associated with the intensity of the XRD spectrum as a function of $2\theta$. Six convolutional layers are employed with kernel sizes of 35, 30, 25, 20, 15, and 10. The stride and filter size are both kept fixed at 1 and 64, respectively, for all layers. ReLU activation is used throughout. Between each convolutional layer, pooling is carried out with a stride of 2. The pool sizes are chosen as 3, 3, 2, 2, 1, and 1. After the final pooling layer, flattening is applied to obtain a one-dimensional vector that is then fed to a fully connected neural network containing two hidden layers with sizes 3,100 and 1,200. The final output layer contains either 140 (255) nodes when dealing with stoichiometric (and non-stoichiometric) phases. Between all layers in the fully connected neural network, a dropout rate of 60% is applied to maintain regularization.

**Training:** The CNN yields one-hot vectors [0, 0, 1, 0, …, 0] where each index represents a reference phase. Accordingly, the loss function is defined as the cross entropy between the true and predicted vectors. An Adam optimizer is utilized to minimize the loss. Training was conducted across 2 epochs using a batch size of 32 and five-fold cross-validation.

**Supplementary Note 3: Profile fitting and subtraction**

**Fitting:** Once a phase has been identified, its diffraction peaks are simulated as described in the Supplementary Note 1. Dynamic time warping (DTW) is carried out between these peaks and the measured spectrum by using the DTW package for Python[1]. As warping aims to match correlated indices between two times series within a given window, it requires a maximum bound to be chosen such that peaks can only be matched with one another if their positions are with $\Delta 2\theta$ of one another. Here, we chose $\Delta 2\theta = 1.5°$ to reflect the extreme magnitude of peak shifts that may arise in experiment, e.g., from strain or off-stoichiometry. Once the indices have been mapped by DTW to provide a fitting along the x-direction ($2\theta$), fitting is performed along the y-direction ($I$). For this, the simulated spectrum is scaled as to minimize the average difference between the intensities of its diffraction peaks and those of the measured spectrum. All peaks with intensities greater than 5% of the maximum peak intensity are identified using the signal processing module from SciPy[2]. The minimal difference is found by sampling 100 scaling constants that range from 0% and 100% of the maximum intensity from the measured spectrum.

**Subtraction:** After the simulated spectrum of the identified phase has been fit, its intensities are subtracted from the measured spectrum. As the fitting is not always perfect, subtraction occasionally produced negative intensities in the resulting spectrum. To avoid any associated issues, all negative values are set to zero.

**Supplementary Note 4: Experimental methods**

**Single-phase samples:** LiF, $TiO_2$ (anatase), MnO, $Mn_2O_3$, $LiMnO_2$, $LiMn_2O_4$, $Li_2TiO_3$, $MnF_2$, and $MnO_2$ powders were supplied by Sigma-Aldrich and used as pristine materials. $Li_2MnO_3$ was synthesized *via* solid-state reaction by mixing stoichiometric $Li_2CO_3$ (Sigma-Aldrich) and $MnO_2$ followed by heating at 800 °C in the air for 12 hours with natural cooling. To sample non-stoichiometry, four disordered rocksalts were synthesized with compositions $Li_{1.2}Mn_{0.4}Ti_{0.4}O_2$, $Li_{1.3}Mn_{0.1}Ti_{0.6}O_2$, $Li_{1.3}Mn_{0.3}Ti_{0.4}O_{1.8}F_{0.2}$, and $Li_{1.25}Mn_{0.45}Ti_{0.3}O_{1.8}F_{0.2}$ by mixing stoichiometric amounts of $Li_2CO_3$, $Mn_2O_3$, $TiO_2$, and LiF and heating at 1000 °C under a flowing argon atmosphere for two hours followed by natural cooling. To reduce the particle size for broadening diffraction peaks, each of LiF, $LiMnO_2$, $MnF_2$, $MnO_2$, and $Li_2TiO_3$ was ball-milled in a high-energy SPEX 800M shaker mill at 1725 rpm for 15-30 minutes.

**Multi-phase mixtures:** To prepare multi-phase mixtures, equivalent masses of two or three pristine materials were mixed with a pestle and mortar for 15 minutes. The following combinations of materials were considered: $MnF_2$ + $TiO_2$, $Mn_2O_3$ + LiF, $TiO_2$ + MnO, $LiMn_2O_4$ + $Li_2MnO_3$, $LiMnO_2$ + $Li_2MnO_3$, $Li_2TiO_3$ + $MnF_2$ + $LiMn_2O_4$, $Mn_2O_3$ + $TiO_2$ + $LiMnO_2$, $Li_2MnO_3$ + MnO + $TiO_2$, $Mn_2O_3$ + $Li_2TiO_3$ + $TiO_2$, $LiMn_2O_4$ + LiF + $Li_2MnO_3$.

**X-ray diffraction measurement:** XRD spectra were measured with a Rigaku MiniFlex 600 using Cu $K_\alpha$ radiation. $2\theta$ was scanned between 10° and 100° using a step size of 0.01°. A scan rate of 3°/minute was applied for all measurements, except when generating noisy signals (in which case a scan rate of 30°/minute was used).

**Supplementary Note 5: Baseline correction and noise filtering**

**Baseline correction:** To identify and subtract the background signal from a given spectrum, we employed the rolling ball algorithm as implemented in the OpenCV package for Python[1]. In one-dimension, this approach may be visualized by imagining the translation of a circle along $2\theta$, with at least one point on the edge of the circle constantly touching the spectrum. Then, any intensity where the circle and spectrum are in contact is assumed to be a part of the background. Here, we choose the radius of the circle as 4° so that diffuse features are attributed to the background while still retaining some allowance for broad peaks. After the background spectrum has been identified, it is subtracted from the measured spectrum.

**Noise filtering:** Noise is removed from measured spectra using an infinite impulse response filter it implemented in the signal processing module from SciPy[2].

**Supplementary Note 6: Training on multi-phase spectra**

**Spectrum simulation:** To provide a comparison with our newly developed approach to phase identification from multi-phase spectra based on an iterative procedure of phase identification and profile subtraction, we designed a separate model based on the work of Lee *et al.*[1] Accordingly, single-phase diffraction spectra were simulated (without data augmentation) from the 140 stoichiometric reference phases spanning the Li-Mn-Ti-O-F composition space. In total, 140,420 and 273,819 spectra were constructed to represent two- and three-phase mixture, respectively. This was done by enumerating all possible combinations of the stoichiometric reference phases, from which diffraction peaks were added together through a linear combination where the coefficients are randomly selected to scale each individual spectrum from 0% to 100% of its initial intensity. Normalization was conducted after performing each linear combination such that maximum intensity is set to 100.

**Training:** A similar CNN architecture was utilized as discussed in the Supplementary Note 2. The only difference lies in the output layer, which was designed to follow a three-hot vector approach whereby each reference phase has three associated indices representing a low, moderate, and high weight fraction. For example, if two phases are present with a low and high weight fraction, then its representation would appear as ([1, 0, 0], [0, 0, 1]). Further details regarding this method can be found in Ref. [1] listed below. Based on this procedure, the output layer of the CNN now contains 420 nodes: 3 weight fractions × 120 reference phases. Training was conducted as previously described.